\documentclass[aps,prl,twocolumn,showpacs,superscriptaddress,groupedaddress]{revtex4}  % for review and submission
\usepackage{graphicx}  % needed for figures
\usepackage{dcolumn}   % needed for some tables
\usepackage{bm}        % for math
\usepackage{amssymb}   % for math
 \usepackage{graphicx}
\usepackage[latin1]{inputenc}
\usepackage{srcltx}

\begin{document}

% The following information is for internal review, please remove them for submission
%\widetext
%\leftline{Version xx as of \today}
%\leftline{Primary authors: M. Gross}
%\leftline{To be submitted to (PRR)}

% the following line is for submission, including submission to the arXiv!!
%\hspace{5.2in} \mbox{Fermilab-Pub-04/xxx-E}

\title{Measuring enhanced optical correlations induced by transmission open channels in a slab geometry}

\author{N.~Verrier}
\author{L.~Depreater}
\author{D. Felbacq}
\author{M.~Gross}
\affiliation{Laboratoire Charles Coulomb  - UMR 5221 CNRS-UM
Universit\'{e} Montpellier Bat 11.
Place Eug\`{e}ne Bataillon
34095 Montpellier }

\date{\today}

\begin{abstract}
Light can be transmitted through disordered media with high efficiency due to the existence of  open channels, but quantitative agreement with theory is  difficult to obtain  because of the large number of channels to control. By measuring, by digital holography, the correlations of the transmitted field, we are able to determine the number of channels involved in transmission. With sample, the number of channels is divided by about 15 with respect to what is expected without sample. This figure is in good agreement with the theoretical prediction.
\end{abstract}

\pacs{42.25.Bs    Wave propagation, transmission and absorption;
05.60.Cd    Classical transport;
02.10.Yn    Matrix theory;
42.40.Ht    Hologram recording and readout methods.
}

\maketitle
%
%42.25.-p    Wave optics
%
%42.25.Dd    Wave propagation in random media
%
%05.40.-a    Fluctuation phenomena, random processes, noise, and Brownian motion

%Averaged correlations $\langle |C|^2\rangle $  and $\langle |C'|^2\rangle $

%\section{\label{sec:level1}First-level heading}
% sections are not used for PRL papers
%TM Eigenvalue:

%
%
%\bigskip
%

%
%\bigskip

Wave propagation in complex media gives rise to a multitude of effects that physicists are trying to understand and analyze both experimentally and theoretically. Some of these effects, such as coherent backscattering were observed \cite{van1985observation, wolf1985weak}, and are well understood. Others, like the 3D Anderson localization of an electromagnetic wave, are a kind of Holy Grail \cite{anderson1958absence} that may exist or not. Other, are under study.  For example, recent  wavefront
shaping and phase recording technologies make possible to focus and image through
strongly scattering media \cite{vellekoop2007focusing,vellekoop2008universal,vellekoop2010exploiting,popoff2010measuring}, or  to get enhanced  transmission  by adjusting the input mode structure to fit with the so-called "open channels" \cite{kim2012maximal,yu2013measuring,popoff2014coherent}. In the best case, the transmission was increased by a factor of 4 with respect to that observed without adjustment \cite{kim2012maximal}, while optimal transmission remains small as compared to the 100 \% expected with open channels \cite{nazarov1994limits}.
%\section{Critique des manips de Popov et Yu}

%Deux méthodes ont été utilisées pour étudier les modes ouverts. La première méthode, mise en oeuvre par Popof et al.  \cite{popoff2014coherent}, consiste à modifier la structure de modes du champ  incidents à l'aide d'un Spacial Light Modulator SLM de manière rendre  maximale la transmission. Une fois ce maximum atteint, on considère que le mode incident correspond à un mode ouvert et que la transmission  maximale atteinte correspond à celle d'un mode ouvert. La seconde méthode,  utilisée  par Yu et al. \cite{yu2013measuring},   consiste à mesurer et à diagonaliser la Transmission Matrix ($T$). Les valeurs propres maximales correspondent alors aux modes ouvert.
Two methods have been used to study this open channel effect.
The first method is to change the mode structure of the incident field with a spatial light modulator (SLM) to maximize the transmission \cite{popoff2014coherent}.
Once the maximum is reached, it is considered that the incident  mode  fits with an open channel, and that the obtained transmission is that of an open channel (ideally 100\%). The second method  is to measure  and to diagonalize the transmission matrix $\textbf{t}$ \cite{popoff2010measuring}. The eigenvectors, whose eigenvalues are maximal, correspond to open channels \cite{kim2012maximal,yu2013measuring}.
%La mise en évidence des modes ouverts en géométrie ouverte (géométrie slab) par ces deux méthodes présente de grande difficultés du fait du grand nombre $N_g$ de modes ou de canaux à contrôler ou à mesurer. En effet, si l'on veut observer des effets  importants il faut que l'épaisseur du milieu $l$ soit grande devant la longueur de transport de l'énergie $l^*$ qui est en pratique de l'ordre de la longueur d'onde $\lambda$. Par ailleurs, si l'on veut  obtenir des effets simples à analyser,  il faut que les dimensions latérales $L$ soient grandes devant l'épaisseur $l$.  On doit donc avoir, $L \gg l \gg l^*$. Avec  $l^* \sim \lambda$, il faut avoir par exemple $\l \sim 10\lambda$, et  $L \sim 100\lambda$.  Pour un échantillon d'aire $L^2$, et avec 2 polarisations, on a $ N_g \sim 4 \pi L^2 /\lambda^2 \sim 10^5$.
%

The study of open channels in an open geometry (slab) by both methods presents great difficulties because of the large number $N_g $ of geometrical modes (or channels) to control or measure. Indeed, the observation of significant effects requires a medium, whose thickness  $ l $ is large compared to the light transport mean free path $ l^* $.
Furthermore, lateral dimensions $ L$  should be large compared with  the thickness $l $ so that the effect can be analyzed easily. So we must have $ L \gg l \gg l^* $, i.e.   for example $ \l \sim 10 l^*  $ and $ L \sim  100 l^* $. On the other hand, it holds $ N_g = 2 \pi S /\lambda^2$ (for 2 polarisations), where $S=L^2$ is the sample area, and $\lambda$ the wavelength. Since $l^* \sim \lambda$ in the best case, we get $N_g \sim 10^5$.
This number is much higher than the number of modes controlled or measured  in experiments  \cite{popoff2014coherent,yu2013measuring}.

Popoff et al. \cite{popoff2014coherent} control only 650 modes. They get an increase in the transmission by a factor  3, but nothing indicates that the optimization process is completed, and that the obtained maximum of transmission  corresponds to an open channel. Yu et al. \cite{yu2013measuring} perform measurements on a base of 21078 modes, and the distribution of the eigenvalues they get is quite far from that predicted by the open channel theory.

\begin{figure}
\begin{center}
  \includegraphics[width=5 cm]{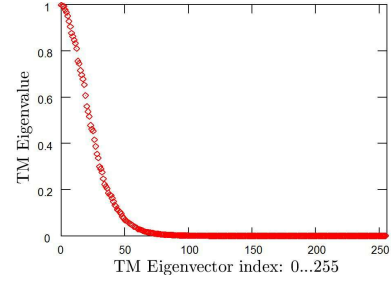}
  \caption{Typical distribution of  the eigenvalues $t_n$ of $\textbf{t t}^\dag$  plotted in decreasing order. Eigenvectors are calculated by using a typical  transmission matrix $[\textbf{t}]$ of size $N\times N$ with $N=256$ obtained by Monte Carlo.  The parameters of the calculation are adjusted to get an average transmission $\sum_{n=1}^{256} t^2_n \simeq 0.1$. }\label{Fig_fig_eigenvalues}
\end{center}
\end{figure}

Rather than preparing open channels  which is a difficult task, we propose
to study the outgoing field, which is directly observable.  This field is distributed over a number $N_2$ of  modes, which is about the number of open channels, that is  $N_g l^*/l$ \cite{pendry1992universality}, this is thus much smaller than the number of geometrical modes $N_g$. For random incoming fields, one expects thus to observe larger correlations for the outgoing field than for the incoming one. %
In the following it is shown that  these correlations can indeed be used to define and measure  $N_2$, that they  provide a mean to highlight the open channel effect by measuring the ratio $N_g/N_2 \gg 1$, and to make quantitative comparison with theory.

 To be more specific, let us consider the eigenvectors $\textbf{U}_n$  of  $\textbf{t}^{\dag} \textbf{t}$ where $\textbf{t}$ is the transmission matrix. These eigenvectors $\textbf{U}_n$, whose eigenvalues are $t_n^2$ (where $t_n>0$ is real)  constitute  a natural  orthonormal  basis of the incoming field $\textbf{E}_1$ with
 $\textbf{U}_n \textbf{.} \textbf{U}_{n'} =\delta(n,n')$.
Similarly, the eigenvectors $\textbf{V}_n$ of $\textbf{t}\textbf{ t}^{\dag}$    constitute an orthonormal basis of the outgoing field $\textbf{E}_2$ with  $\textbf{V}_n \textbf{.} \textbf{V}_{n'} =\delta(n,n')$. $\textbf{U}_n$ and $\textbf{V}_n$ are related together  by  $ \textbf{t} \textbf{U}_n = t_n \textbf{V}_n$.
We calculated, by Monte Carlo, a  typical transmission matrix $\textbf{t}$  obtained for  $N_g=256$ modes in the random matrix theory ideal case, and we calculated the eigenvector $\textbf{U}_n$ and the eigenvalue $t^2_n$. Like Choi et al. \cite{choi2011transmission}, we have plotted  on Fig.\ref{Fig_fig_eigenvalues}   the  eigenvalues $t_n^2$ in decreasing order  from   $t^2_n=1$ for $n=1$ down to $t^2_n \simeq 0$ for $n=256$.
The calculation was made for $l=9 l^*$ to get an average transmission for energy
%\begin{equation}\label{Eq_T}
    $ T  = (1/N) \sum_{n=1}^{N} t_n^2$
%\end{equation}
 close to 0.1. The open channels correspond here to the first eigenvectors, with $t^2_n \simeq 1$, while the outgoing channels correspond to the channels, whose transmission $t_n^2$ is noticeable ($n= 1$ to about 40). $\textbf{E}_1$ is distributed over all  modes  $\textbf{U}_n$ with $n=1$ to 256, while the outgoing field $\textbf{E}_2$ is mainly distributed over the so-called ''outgoing modes''  $\textbf{V}_n$ with $n=1$ to about 40.

Let us consider a  typical input field $\textbf{E}_1=\sum_{n=1}^{N_g} c_n \textbf{U}_n$, where $c_n$ are independent identically distributed Gaussian complex random numbers.
The  outgoing field is $\textbf{E}_2=\textbf{t} \textbf{E}_1= \sum_{n=1}^{N} c_n t_n \textbf{V}_n$. The incoming and  outgoing  energies are respectively
 $|\textbf{E}_1|^2=\sum_{n=1}^{N_g}  |c_n|^2 $
and
 $|\textbf{E}_2|^2=\sum_{n=1}^{N_g}t_n^2 |c_n|^2$.
The Monte Carlo calculation gives a transmission efficiency ${|\textbf{E}_2|^2}/{ |\textbf{E}_1|^2}$ close to $T=0.1$.
 If $\textbf{E}_2$ is  re-injected into the sample by using a phase conjugaison mirror, we get the twice transmitted field (i.e. the outgoing field that is back transmitted through the sample) $\textbf{E}_3= \sum_{n=1}^{N_g} c_n t^2_n \textbf{U}_n  $, whose energy is $|\textbf{E}_3|^2=  \sum_{n=1}^{N_g} t_n^4 |c_n|^2$.  We can define the back transmission $T_B$ of the outgoing field by $T_B$=${|\textbf{E}_3|^2}/{ |\textbf{E}_2|^2}$. We got $T_B$=2/3  as expected  \cite{nazarov1994limits,vellekoop2008universal}.

In this letter, we propose to study  the "equivalent numbers  of  modes"  $N_1$ and $N_2$ of the incoming ($\textbf{E}_1$) and outgoing field ($\textbf{E}_2$). Since  $\textbf{E}_1$ is uniformly distributed over all the  $N_g$ geometrical modes $\textbf{U}_n$, one has $N_1$=$N_g$.  On the other hand, it holds  $N_2\ll N_g $ since $\textbf{E}_2$ is distributed over a much smaller number of outgoing modes $\textbf{V}_n$. The "equivalent number  of  modes" $N_2$ corresponds here  to the ''width''  of the curve $t^2_n$ (i.e.  $N_2\simeq$40 for the example plotted on Fig.\ref{Fig_fig_eigenvalues}).

In order to define $N_2$ more precisely, we considered two uncorrelated random incoming  fields $\textbf{E}_1=\sum_{n=1}^{N_g} c_n \textbf{U}_n$ and $\textbf{E'}_1=\sum_{n=1}^{N_g} c'_n \textbf{U}_n$ having the same energy $|\textbf{E}_1|^2= |\textbf{E'}_1|^2$, and the corresponding outgoing  fields $\textbf{E}_2=\sum_{n=1}^{N_g} c_n t_n \textbf{V}_n$ and $\textbf{E'}_2=\sum_{n=1}^{N_g} c'_n t_n \textbf{V}_n$. One can then calculate the normalized residual correlation $C_1$ and $C_2$  for the incoming and outgoing fields.
\begin{eqnarray}\label{eq_corr_C_2}
  C_1  &=&  \frac{ \textbf{E}_1 \textbf{.}\textbf{ E'}_1^*}{  |\textbf{E}_1|^2}= \frac{\sum_{n=1}^{N_g} c_n c'^*_n} {\sum_{n=1}^{N_g} |c_n|^2 }\\
\nonumber C_2  &=&   \frac{\textbf{E}_2 \textbf{.}\textbf{ E'}_2^*}{ |\textbf{E}_2|^2}=\frac{\sum_{n=1}^{N_g} c_n c'^*_n t^2_n } {\sum_{n=1}^{N_g} |c_n|^2 t_n^2}
\end{eqnarray}
We calculated the statistical averages $\langle  |C_1|^2 \rangle$ by Monte Carlo and got  $\langle  |C_1|^2 \rangle = 1/N_g=1/N_1$. This result can be proved rigorously (see the supplementary material).
This suggest to define the equivalent number of mode   $N_2$  by   $\langle  |C_2|^2 \rangle=1/N_2$.
We have then calculated typical transmission matrices $\textbf{t}$ for  $N_g$=256, 512 and 1024 and for $l/l^*$=1 to 40. For each $\textbf{t}$ we have calculated  $\langle  |C_2|^2 \rangle$ (and thus $N_2$) by Monte Carlo.
%,  by averaging $|C_1|^2$ and   $|C_2|^2$ over $10^4$ couples of random uncorrelated incoming fields $\textbf{E}_1$ and $\textbf{E'}_1$.
For $l/l^*\gg 1$, we got the simple empirical formula: % $\langle  |C_1|^2 \rangle \simeq 1/N$  as expected, and $\langle  |C_2|^2 \rangle \simeq  1/N_2$,  with
\begin{eqnarray}\label{Eq_N_2_theoretical_ideal}
  N_2 &=&(3/2) T N_g
\end{eqnarray}
%where $T$ %given by Eq.\ref{Eq_T}
%is the average transmission.
%
The number of outgoing mode $N_2$ is thus $3/2 \times$ larger than number of open channels ($T N_g$).

The key idea of this work is that  $\langle |C_2|^2\rangle$ and $N_2$ can be easily  measured  experimentally, with a time varying random incoming field, which is uncorrelated at times $t$ and $t'$. Indeed, one can get the correlation   $C_2$  from the measured outgoing field $E_2(x,y,t,p)$ (where  $x,y$ are the transverse coordinates, $t$ the time, and $p=1,2$ the polarisation) by:
\begin{eqnarray}\label{eq1_corr_C_1_C_2_bis}
%  C_1  &=&  = \frac{\sum_{x,y,p} E_1 (x,y,t_1,p) E_1^*(x,y,t_2,p)} {\sum_{x,y,p} |E_1(x,y,t_1,p)|^2 }\\
 C_2(t,t')  &=&   \frac{\sum_{x,y,p} E_2(x,y,t,p) E_2^*(x,y,t',p)} {\sum_{x,y,p} |E_2(x,y,t,p)|^2 }
\end{eqnarray}
%
%where $E_2$ is the  outgoing field measured at point $x,y$, at times $t_1$ and $t_2$ and for polarisation $p=1,2$ .
%
The importance of open channels can be directly demonstrated by   measuring $N_2=1/\langle |C_2|^2\rangle$ and by verifying that $ N_2/N_g$$\ll$1. Furthermore, a   quantitative comparison with our  empirical model can be made by verifying   Eq.\ref{Eq_N_2_theoretical_ideal} experimentally.

\begin{figure}[]
\begin{center}
  \includegraphics[width=6cm]{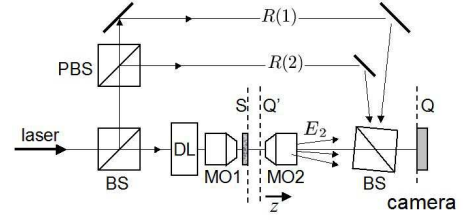}
  \caption{Scheme of the experimental setup.  $R(1)$, $R(2)$:  reference fields of polarisation $p$=1 and $p$=2;  BS, PBS, beam splitter and  polarized beam splitter;  DL: diffusing liquid; MO1 and MO2: microscope objectives; Q, Q': camera plane, and camera  conjugate plane with MO2; S: sample and sample outgoing plane.}\label{Fig_Fig_setup}
\end{center}
\end{figure}

%
%
%
%\begin{figure}[]
%\begin{center}
%     % Requires \usepackage{graphicx}
%  \includegraphics[width=6cm]{images/Fig_setup3}
%  \caption{Experimental setup.  ${E}_{1}$: incoming field; $E_2$: outgoing field;  $R(1)$, $R(2)$:  reference fields of polarisation $p$=1 and $p$=2;  PBS1, PBS2: polarized beam splitters; $\lambda$/2: half wave plate to control  $E_1$, $R(1)$ and $R(2)$ respective power; BS: beam splitter; M: mirror; DL: diffusing liquid; MO1 and MO2: microscope objectives; Q, Q': camera plane, and camera conjugate plane with respect to MO2; S: sample outgoing plane.}\label{Fig_Fig_setup}
%\end{center}
%\end{figure}
%
%

Figure \ref{Fig_Fig_setup} shows the scheme of  holographic setup.   The  sample S is a ZnO powder slab with thickness $l=22\ \mu\rm m\pm7\ \mu\rm m$ deposited on a microscope cover slide. In order to maximize the collection of both input and output modes, the sample is positioned between two microscope objectives: MO1 (${\rm NA=0.9}$ air, x60) in the powder side, and MO2 (${\rm NA=1.4}$ oil, x60) in the cover slide side. A  tank  (DL) filled with viscous diffusing liquid  is positioned in front of MO1 to randomize the illumination structure in both time and space.  The outgoing fields $E_2(x,y,t,p)$ interfere with two orthogonally polarized reference beams $R(p)$ whose off axis angles are slightly different. It is then possible to extract  $E_2(x,y,t,p)$  from the  interference pattern $I=\sum_{p=1,2}|R(p)+E_2(p)|^2$ that is recorded by the camera. This calculation involves  2 frames holograms $H=I_n-I_{n+1}$ where $I_n$ and $I_{n+1}$ are successive camera frames. See supplementary to get details on experiment and on data analysis.
It should be noted that the experimental  tradeoffs  (i.e. diffusing liquid randomized illumination instead of SLM and  two frame holograms) were made in order  to minimize the correlation bias within the incoming fields $E_1$. In particular, two frames holograms  minimize bias due to  ballistic light that travel through the diffusing tank DL.

\begin{figure}
  \includegraphics[width=4 cm]{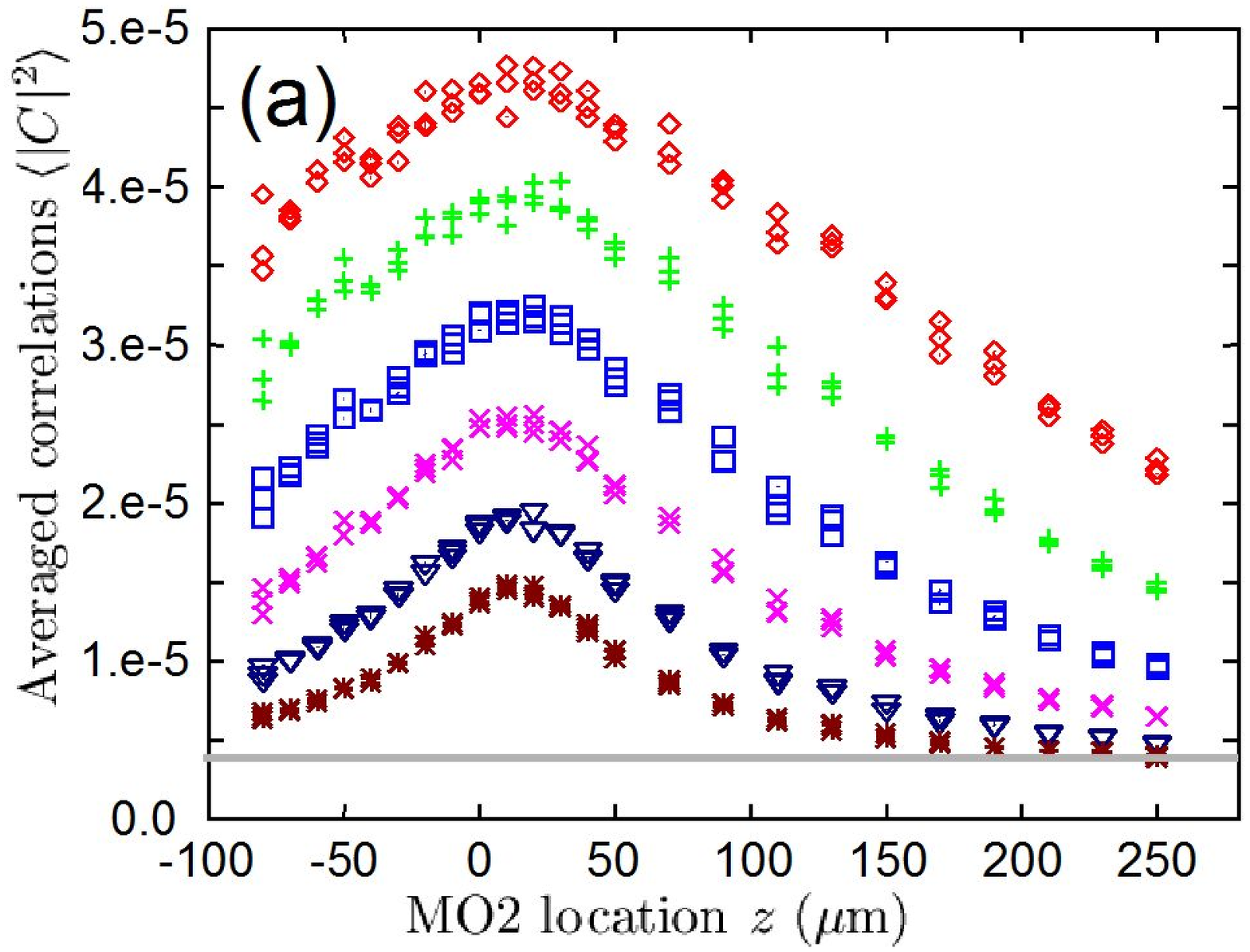}
  \includegraphics[width=4 cm]{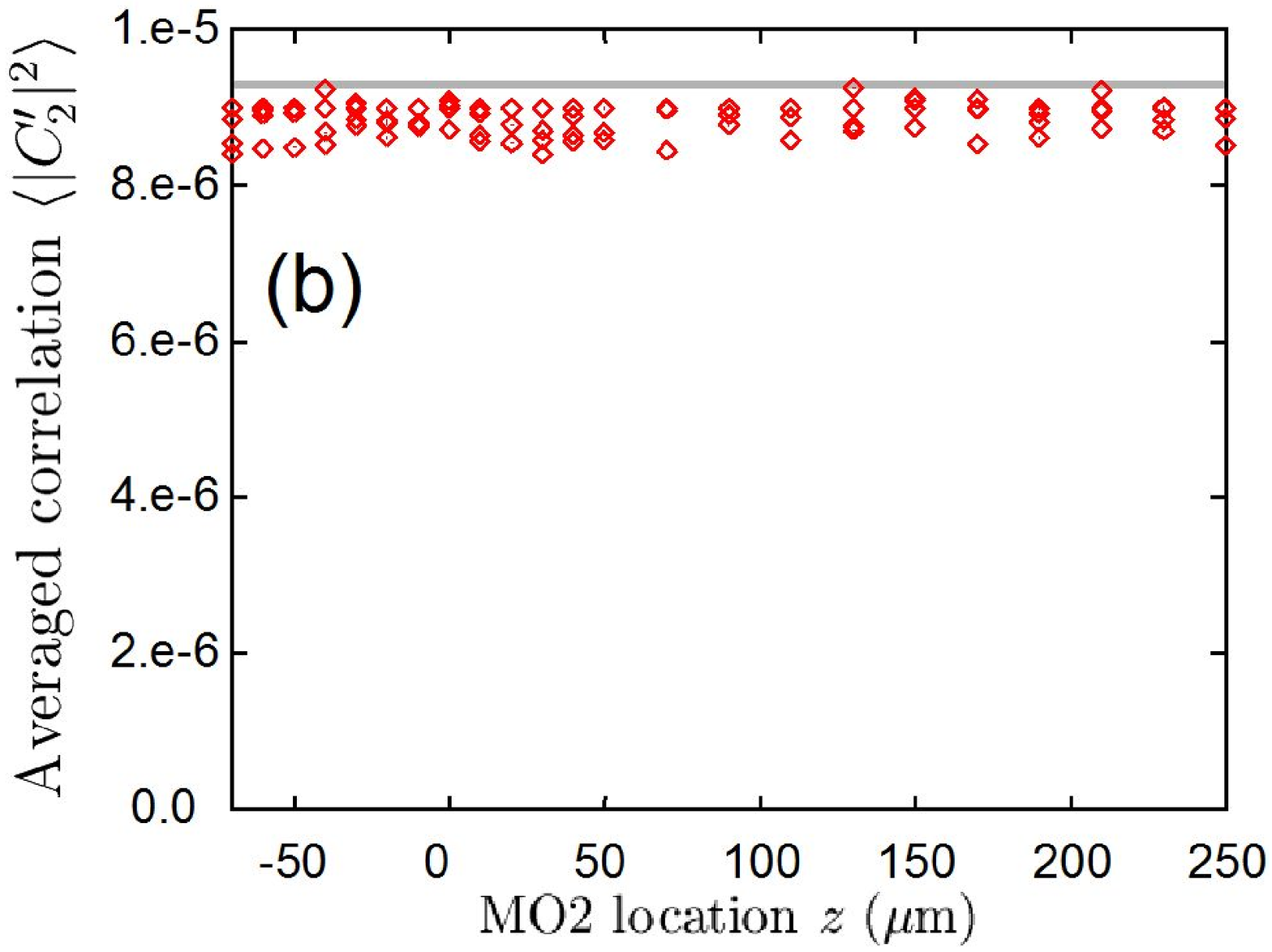}\\
  \caption{ (a) Averaged correlation $\langle |C|^2 \rangle $  measured with  ZnO   plotted as function of  MO2 position $z$. Calculation is made with $M \times M$ pixels with $M=1024$ (red diamonds), 362 (green crosses), 256  (bleue rectangles), 181 (purple crosses), 128 (dark blue triangles) and  90 (brown stars).  Solid grey lines is  $1/N_g$=3.83e-6. (b) Averaged correlation $\langle |C'|^2 \rangle $ measured without  ZnO. Solid grey line is     $1/N'_g$=9.286e-6. }\label{fig__curves}
\end{figure}

To have an extra control parameter, the MO2 position along $ z $ is adjusted by a motorized micrometer device. For $ z \simeq $ 0, the camera's conjugated plane Q' coincides with the exit plane S of the sample, and plane S  is  on focus on the camera. This configuration maximizes the open channel effect. %
When $ | z | \ne 0 $ increases, the  measurement plane Q' no longer coincides with  S and  $\langle |C_2|^2 \rangle $  decreases.
Indeed, if  Q' $\ne $ S, the correlation $ C_2 $ is calculated without being able to define precisely the sample selected zone where the outgoing field is measured:  points outside of the zone contribute to the signal, and points inside the zone contribute less since the detection angle is  lower \footnote{The angle effect was calculated by Goetchy et al. \cite{goetschy2013filtering}}.
Controlling $z$  provides thus an experimental mean to decrease and even to suppress the  open channel effect. This is expected to occur when  $z$ becomes much larger than the sample transverse size $L$. In that case,  $\langle |C_2|^2 \rangle $  must be close to $\langle |C_1|^2 \rangle=1/N_g $. Data are acquired by moving MO2 from $ z $=$ \textrm{Q'S} $=-80 to 250 $ \mu$m by step $\Delta z$=10 or 20 $\mu$m. For each position $z$ of MO2, 150 frames $I$ are recorded and  $\langle |C_2|^2 \rangle $ is calculated with 75 holograms $H$.

In order to vary the sample transverse size  $L$, the data recorded by the camera ($1340\times 1024$ pixels) are cropped to $M \times M$  with $L=77~\mu$m for $M=1024$. The average correlation $\langle |C_2|^2 \rangle $  is multiplied by $M^2/1024^2$ to compensate for area variation.  %for different transverse size  of the selected zone.  This is made by cropping the  data over $x$, and $y$.
%The cropped zones are represented by  the  dotted squares in Fig.\ref{Fig_ima_pupil} (c). This factor is equal to 1 for  (i.e. $L=77~\mu$m).
%%
This provides a second experimental mean to decrease  the effect of open channels when the transverse size $L$ of the sample becomes smaller than its thickness $l$, since, in that case, the transverse geometrical losses becomes large. Conversely, by extrapolating $L$ toward infinite, one can get an estimate of the maximum enhancement of correlation  that is expected for a given thickness $l$ of the sample.

We have verified experimentally that changing $z$ and $M$ does not modify the detected energy per unit of area.
We have plotted on Fig. \ref{fig__curves} (a) $M^2 \langle |C_2|^2 \rangle /1024^2$   as a function of the MO2 position $z$ for $M=1024$ (red diamonds), 362 (green crosses), 256  (bleue rectangles), 181 (purple crosses), 128 (dark blue triangles) and  90 (brown stars).
As expected $ \langle |C_2|^2 \rangle $ is maximum for $z\simeq 0$, and decreases with $|z|$.   When the sample size $M$ decreases, the geometrical losses increase, and  $M^2 \langle |C_2|^2 \rangle /1024^2$ decreases too. Moreover, the peak of correlation becomes narrower, and the correlation reaches a plateau for small $M$ and large $z$ (\emph{e.g.} $M=90$ and $z=250~\mu$m). This plateau is reached when geometrical losses are large, that is when   $z\gg l,L$.

We compared this plateau with the correlation $ \langle |C_1|^2 \rangle =1/N_g$ that is expected without open channels. The usual relation that defines the number of geometrical mode $N_g$ should be  corrected to account for the refractive index $n$ of the medium where the output field is  collected (factor $\times n^2$), and for the detection solid angle (factor $\times [\textrm{NA}]^2/n^2$) where NA is the numerical aperture of the detection objective. One gets thus:
\begin{equation}\label{Eq_N_g}
     N_g=2\pi \textrm{[\textrm{NA}]}^2 L^2/\lambda^2
\end{equation}
where $L$ is the transverse size of the sample and $\lambda$ the wavelength in vacuum.
For NA=1.4 and $L$=77 $\mu$m, we get $1/Ng$=3.83e-6 (solid grey line on Fig.\ref{fig__curves}(a)) in good agreement with the plateau of correlation curves $M^2 \langle |C_2|^2 \rangle /1024^2$. This agreement validates our correlation-based  theoretical approach and proves that  we have correctly evaluated the number of geometrical modes by Eq.\ref{Eq_N_g} .

We conducted a control experiment without  ZnO sample and we calculated the correlation  $\langle |C'_2|^2 \rangle $ versus $z$ for $M=1024$. As seen on  Fig.\ref{fig__curves}  (b), $\langle |C'_2|^2 \rangle $  is much lower than with the sample, and does not depend on $ z $ position. Moreover, $\langle |C'_2|^2 \rangle \simeq$8.8e-6 is close to the correlation  that is expected without open  channels  $ \langle |C'_1|^2 \rangle =1/N'_g$=9.3e-6 (see solid grey line on  Fig.\ref{fig__curves}(b)), where $N'g$ is calculated by Eq.\ref{Eq_N_g} with  MO1 numerical aperture  NA=0.9 (see supplementary).
%%
%Indeed, the signal collection angle is limited by MO1. %, and therefore the relevant  numerical aperture is that of MO1.
%This point is illustrated by Fig.\ref{Fig_ima_pupil} (d) that displays the hologram $ |\tilde H|^2 $ we got without sample. $ |\tilde H|^2 $ exhibits four bright circular zones that are smaller in diameter  than with the sample (Fig.\ref{Fig_ima_pupil}(a)). These circles correspond to the MO1 pupil located in the plane P' that appears sharp in the  plane P, because MO1 and MO2 form an afocal optical system.

\begin{figure}
  \includegraphics[width=5 cm]{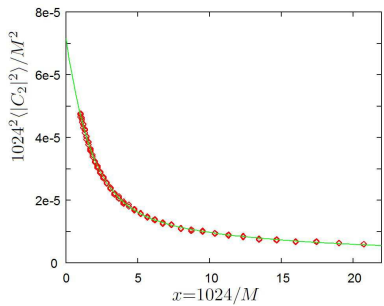}
  \caption{Maximum   averaged correlation $1024^2 \langle |C_2|^2 \rangle /M^2$  according to  $x= 1024/M$. Measured point are red diamonds. Solid green line is $ a_1 e^{ -x/b_1} + a_2 e^{ -x/b_3} + a_3$, with  $a_1$=5.19e-5, $a_2$=1.59e-5, $b_1$=1.7,  $b_2$=10.0 and  $a_3$=3.67e-6$\simeq$1/$N_g$.  }\label{fig__curves_L}
\end{figure}

To calculate the correlation that is expected without lateral geometrical losses of you slab sample, %(i.e. for $L=\infty$)
we have plotted on Fig. \ref{fig__curves_L} (red diamonds),
the maximum of $M^2 \langle |C_2|^2 \rangle /1024^2$ obtained during
the $z$ scan, as a function of $x=1024/M$. We extrapolated our results  to  $L=\infty$ by fitting the  data % have been fitted
with a sum of 2 decreasing exponentials (see solid green line on Fig.\ref{fig__curves_L}).
%
%   Fit  is plotted  with solid green curve. To get  better accuracy, $a_3$ is obtained by fitting data averaged from $z$=50 to 250 $\mu$m.
%%
Extrapolation to $L$=$\infty$ (i.e. $x=0$) yields  $M^2 \langle |C_2|^2 \rangle /1024^2$=7.15e-5. Half maximum is reached for $x$=1.73 i.e. $L$=50 $\mu$m (about twice the thickness $l$).
The importance of open channels that corresponds to $N_g/N_2\gg1$  is here clearly demonstrated, since   we got $N_g/N_2$=12.3 without extrapolation for $L=77~ \mu$m, and 18.6 after extrapolation to $L=\infty$.

The  $N_g/N_2$ measured ratio must {\it a priori} be corrected by the collection angle of the objectives MO1 and MO2 \cite{goetschy2013filtering}. MO2 losses yield a geometrical loss factor $[\textrm{NA}]^2/n^2$=0.85 that affects $N_g$ (see Eq.\ref{Eq_N_g}). %, the MO2 loss factor .
Monte Carlo calculations \footnote{similar to the ones made to get Eq. \ref{Eq_N_2_theoretical_ideal}}
show that they affect  $N_2$ by roughly the same amount when  $N_g/N_2\gg 1$. The ratio $N_g/N_2$ is thus not modified by the MO2  losses.
On the other hand, the  MO1  losses  decrease the number of incoming geometrical modes $N_1=N'_g$ by a factor  0.9$^2$/1.52$^2$=0.35. This factor was accounted for to get the $1/N'g$ solid grey line displayed on Fig.\ref{fig__curves} (b). Calculations show that this decrease of $N_1$  by 0.35 yields a small  decrease of  $N_2$ ($\simeq 0.9$) as noticed by Kim et al.  \cite{kim2013relation}. The corrected extrapolated ratio is thus $N_g/N_2$=16.74.

To perform a quantitative comparison of our results with Eq.\ref{Eq_N_2_theoretical_ideal} empirical formula, we
measured the transmission $T$. The  ratio of the signal energy ($\sum |E_2|^2$) detected with and without sample is  1/30 $\pm$ 20 \% \footnote{This measurement can be  affected by experimental  drifts since the energy measurements made with and without sample  are made at different days.}.
Taking into account the MO2  losses ($\times$0.85) \footnote{Without sample, all the light collected by MO1 is collected by MO2. With sample, 15\% of the outgoing light  is scattered out of the MO2 collecting angle. },  we get $T=(1/30)/0.85\simeq 1/25$.
The ratio  that is expected from  Eq.\ref{Eq_N_2_theoretical_ideal} is thus
$N_g/N_2$=2/3$T$=16.6 $\pm$ 20 \%, in good agreement with the corrected  extrapolated experimental results 16.74. Although this agreement is already extremely encouraging,   the proposed approach can still be improved to provide better quantitative results.  A more precise measurement of the sample thickness $l$,  and  average transmission $T$  should be made. A setup using a high NA oil objective for MO1 yielding roughly the same number of geometrical modes for input and output (i.e. $N_g \simeq N'_g$) would also give a better highlight of the open channel effect.

More fundamentally, the shapes of  Fig.\ref{fig__curves} (a) network curves, and of  Fig.\ref{fig__curves_L} curve give informations that were exploited not. Indeed,  these curves characterize  the dependance of  $\langle|C^2|\rangle$  with the MO2 defocus $z$, the sample transverse size $L$ and the sample thickness $l$, i.e. with the slab geometry. In order to exploit these  informations a theory that accounts for the  slab geometry should be developed. We hope that the present results will encourage such development. We acknowledge  ANR-2011-BS04-017-04 grant for funding.

%
%We must nevertheless mentioned that our experiment is not  quantitative yet, because we have consider here only one sample, and because $T$ has not been measured with enough accuracy. The quantitative comparison with  Eq.\ref{Eq_N_2_theoretical_ideal} will be done in a further work.
%
%
%
%It will be then necessary to account for the effect of
%the geometrical losses  of MO2 ($(1.4/1.52)^2$) and  MO1($(0.9/1.52)^2$). Numerical simulation  shows that for $N_2/N_g\sim$15,  MO2 losses do not affect the ratio $N_2/N_g$, while MO1 losses affects the ratio by about 10\%. For the present work, the expected ratio becomes then 1/18.2 in better agreement with experiment.
%
%It could be also interesting to study the effect of geometrical losses due to the finite transverse size $L$ of our slab sample by developing  a theory able to calculate how  $1024^2 \langle |C_2|^2 \rangle/M^2$ varies with $z$ and $M$. It will be then possible to predict  the shape of the curves of Fig.\ref{fig__curves}(a) and Fig.\ref{fig__curves_L}.
%

\bibliographystyle{unsrt}

%\bibliography{biblio}

\begin{thebibliography}{}

\end{thebibliography}


\begin{thebibliography}{10}


\bibitem{van1985observation}
M.P. Van~Albada and A. Lagendijk.
%\newblock Observation of weak localization of light in a random medium.
\newblock {\em Phys. Rev. Lett.}, \textbf{55}, 2692 (1985).

\bibitem{wolf1985weak}
P.E. Wolf and G. Maret.
%\newblock Weak localization and coherent backscattering of photons in
%  disordered media.
\newblock {\em Phys. Rev. Lett.} \textbf{55}, 2696, (1985).

\bibitem{anderson1958absence}
Philip~W Anderson.
%\newblock Absence of diffusion in certain random lattices.
\newblock {\em Phys. Rev.} \textbf{109}, 1492  (1958).

\bibitem{vellekoop2007focusing}
I.M.  Vellekoop et al; %\overleftarrow{}and A.P. Mosk.
%\newblock Focusing coherent light through opaque strongly scattering media.
\newblock {\em Opt. Lett.} \textbf{32} 2309  (2007).

\bibitem{vellekoop2008universal}
IM~Vellekoop et al. %and AP~Mosk.
%\newblock Universal optimal transmission of light through disordered materials.
\newblock {\em Phys. Rev. Lett.} \textbf{101} 120601 (2008).

\bibitem{vellekoop2010exploiting}
I.M. Vellekoop et al. %, A. Lagendijk, and A.P.Mosk.
%\newblock Exploiting disorder for perfect focusing.
\newblock {\em Nature Photon.} \textbf{4} 320 (2010).

\bibitem{popoff2010measuring}
S.M. Popoff et al. %, G. Lerosey, R. Carminati, M.Fink, A.C. Boccara, and S. Gigan.
%\newblock Measuring the transmission matrix in optics: an approach to the study
 % and control of light propagation in disordered media.
\newblock {\em Phys. Rev. Lett.}  \textbf{104} 100601  (2010).

\bibitem{kim2012maximal}
M. Kim et al. %, Y. Choi, C. Yoon, W. Choi, J. Kim, Q.H.
  % Park, and W. Choi.
%\newblock Maximal energy transport through disordered media with the
%  implementation of transmission eigenchannels.
\newblock {\em Nature Photon.}, \textbf{6} 581 (2012).

\bibitem{yu2013measuring}
H. Yu et al. %, T.R. Hillman, W. Choi, J.O. Lee, M.S. Feld,
 % R.R. Dasari, and Y. Park.
%\newblock Measuring large optical transmission matrices of disordered media.
\newblock {\em Phys. Rev. Lett.} \textbf{111} 153902 (2013).

\bibitem{popoff2014coherent}
S.M. Popoff,  A. Goetschy, S.F. Liew, A.D. Stone, and H. Cao.
%\newblock Coherent control of total transmission of light through disordered
 % media.
\newblock {\em Phys. Rev. Lett.} \textbf{112} 133903  (2014).

\bibitem{nazarov1994limits}
Y.V. Nazarov.
%\newblock Limits of universality in disordered conductors.
\newblock {\em Phys. Rev. Lett.} \textbf{73} 134 (1994).

\bibitem{pendry1992universality}
J.B. Pendry et al. %, A. MacKinnon, and P.J. Roberts.
%\newblock Universality classes and fluctuations in disordered systems.
\newblock In {\em Proc. Roy. Soc. A} \textbf{437},  67 (1992).

\bibitem{choi2011transmission}
W. Choi, A. P. Mosk, Q.~Han Park and  W. Choi.
%\newblock Transmission eigenchannels in a disordered medium.
\newblock {\em Phys. Rev. B} \textbf{83} 134207 (2011).

%\bibitem{cuche2000spatial}
%E. Cuche, P. Marquet, and C. Depeursinge.
%\newblock Spatial filtering for zero-order and twin-image elimination in
%  digital off-axis holography.
%\newblock {\em Appl. Opt.} \textbf{39} 4070 (2000).

\bibitem{goetschy2013filtering}
A. Goetschy and A.D. Stone.
%\newblock Filtering random matrices: the effect of incomplete channel control
 % in multiple scattering.
\newblock {\em Phys. Rev. Lett.},  \textbf{111} 063901 (2013).

\bibitem{kim2013relation}
M.Kim et al. %, W. Choi, C. Yoon, G.H. Kim, and W. Choi.
%\newblock Relation between transmission eigenchannels and single-channel
%  optimizing modes in a disordered medium.
\newblock {\em Opt. Lett.}, \textbf{38} 2994 (2013).

\end{thebibliography}

\end{document}

% The following information is for internal review, please remove them for submission

% the following line is for submission, including submission to the arXiv!!
%\hspace{5.2in} \mbox{Fermilab-Pub-04/xxx-E}

\title{Measuring enhanced optical correlation induced by transmission  open channels in slab geometry (supplementary)}

\author{N.~Verrier}
\author{L.~Depreater}
\author{D. Felbacq}
\author{M.~Gross}
\affiliation{Laboratoire Charles Coulomb  - UMR 5221 CNRS-UM
Universit\'{e} Montpellier  Bat 11.
Place Eug\`{e}ne Bataillon
34095 Montpellier }

%\date{\today}

\pacs{42.25.Bs    Wave propagation, transmission and absorption;
05.60.Cd    Classical transport;
02.10.Yn    Matrix theory;
42.40.Ht    Hologram recording and readout methods.
}

\maketitle

\section{Analytical calculation of  $\left< |C_1|^2\right >$ }

Consider two  random gaussien  vectors $\textbf{E}_1=\sum_{n=1}^{N_g} c_n$ and $\textbf{E'}_1=\sum_{n=1}^{N_g} c'_n$,  where  $c_n$ and $c'_n$ are random  complex gaussien variable  with $\langle c_n\rangle = \langle c'_n\rangle =0$  and   $\langle |c_n|^2 \rangle = \langle |c'_n|^2 \rangle =1$.
Let us consider the correlation $C_1$:
\begin{eqnarray}
C_1= \frac{\textbf{E}_1 \textbf{.} \textbf{E'}_1}{\left |\textbf{E}_1\right |^2}=\frac{1}{\left |\textbf{E}_1\right |^2}\sum_{n=1}^{N_g} c_n c'^*_n
\end{eqnarray}
Let us calculate  $\left< |C_1|^2\right >$ with the Law of large Numbers. We get:
\begin{eqnarray}
  \left< |C_1|^2\right > &=& \frac{1}{N_g^2}\left<  \left|\sum_{n=1}^{N_g} c_n c'^*_n \right|^2 \right > \\
\nonumber   &=&  \frac{1}{N_g^2}\left<  \sum_{n=1}^{N_g} \sum_{p=1}^{N_g} c_n c'^*_n  c^*_p c'_p \right>
  % &=& \frac{1}{N_g^2}\sum _p \left<  |c_p|^2 \right >\left<|c'_p|^2 \right >=\frac{1}{N}
\end{eqnarray}
 Here, the  $n \ne p$ terms do not contribute to $ \left< |C_1|^2\right > $, because  $c_n$, $c'_n$, $c_p$  and  $c'_p$ are statistically independent.  We get thus:
\begin{eqnarray}
  \left< |C_1|^2\right >  &=& \frac{1}{N_g^2} \sum_{n=1}^{N_g} \left<  |c_n|^2  |c'_n|^2   \right >
\end{eqnarray}
Since  $|c_n|^2$ and   $|c'n|^2 $ are also statistically independent, we get:
\begin{eqnarray}
  \left< |C_1|^2\right >  = \frac{1}{N_g^2} \sum_{n=1}^{N_g} \left<  |c_n|^2  \right >  \left< |c'_n|^2   \right> =\frac{1}{N_g}
\end{eqnarray}
This proves  rigorously the result obtained by Monte Carlo for $\left< |C_1|^2\right >$ .

\section{Experimental setup and data analysis  details }

%
%\begin{figure}[]
%\begin{center}
%     % Requires \usepackage{graphicx}
%  \includegraphics[width=6cm]{images/Fig_setup4}
%  \caption{Scheme of the experimental setup.  $R(1)$, $R(2)$:  reference fields of polarisation $p$=1 and $p$=2;  BS, PBS, beam splitter and  polarized beam splitter;  DL: diffusing liquid; MO1 and MO2: microscope objectives; Q, Q': camera plane, and camera  conjugate plane with MO2; S: sample and sample outgoing plane.}\label{Fig_Fig_setup}
%\end{center}
%\end{figure}
%

\begin{figure}[]
\begin{center}
  \includegraphics[width=6cm]{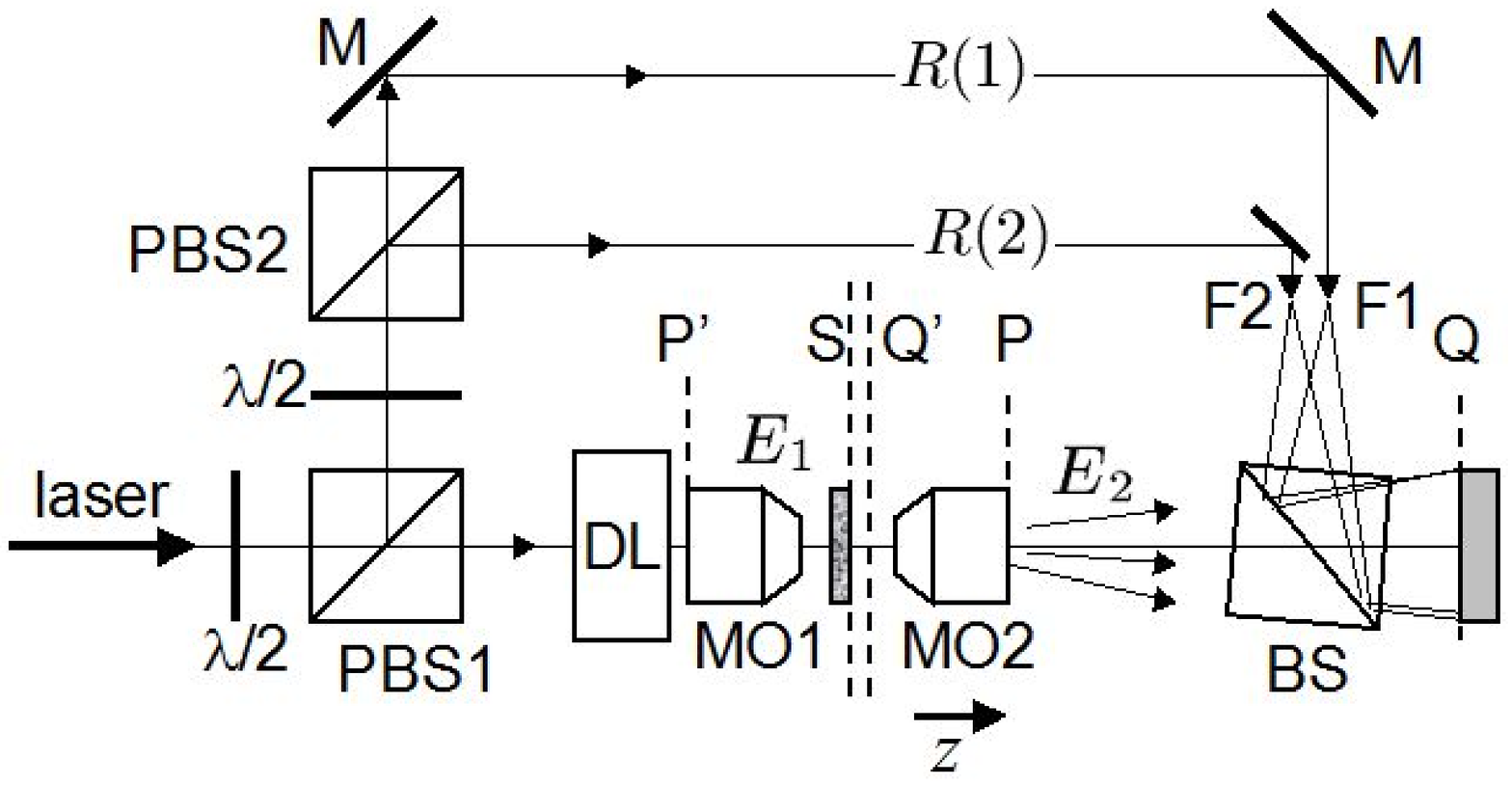}
  \caption{Experimental setup.  ${E}_{1}$: incoming field; $E_2$: outgoing field;  $R(1)$, $R(2)$:  reference fields of polarisation $p$=1 and $p$=2;  PBS1, PBS2: polarized beam splitters; $\lambda$/2: half wave plate to control  $E_1$, $R(1)$ and $R(2)$ respective power; BS: beam splitter; M: mirror; DL: diffusing liquid; MO1 and MO2: microscope objectives; Q, Q': camera plane, and camera conjugate plane with respect to MO2;  S: sample outgoing plane. P' and P: MO1 and MO2 pupil planes.}\label{Fig_Fig_setup}
\end{center}
\end{figure}

Figure \ref{Fig_Fig_setup} shows the setup we used to  study the  open channels  by measuring $E_2(x,y,t,p)$  and  by calculating $C_2$ by :
\begin{eqnarray}\label{eq1_corr_C_1_C_2_bis_bis}
%  C_1  &=&  = \frac{\sum_{x,y,p} E_1 (x,y,t_1,p) E_1^*(x,y,t_2,p)} {\sum_{x,y,p} |E_1(x,y,t_1,p)|^2 }\\
 C_2(t,t')  &=&   \frac{\sum_{x,y,p} E_2(x,y,t,p) E_2^*(x,y,t',p)} {\sum_{x,y,p} |E_2(x,y,t,p)|^2 }
\end{eqnarray}
The setup consists of a Mach-Zehnder off-axis interferometer with two orthogonally polarized reference beams. %$R(p)$  with $p=1,2$ is polarization.
The light emitted by a $\lambda=532\ \rm nm$, 70 mW laser is split into a reference and an object field using a polarizing beam splitter (PBS1). The studied sample is a ZnO powder slab with thickness $l=22\ \mu\rm m\pm7\ \mu\rm m$ deposited on a microscope cover slide. In order to maximize the collection of both input and output modes, the sample is positioned between two microscope objectives: MO1 (${\rm NA=0.9}$ air, x60) in the powder side, and MO2 (${\rm NA=1.4}$ oil, x60) in the cover slide side.

A  tank (1.5 cm thick) filled with viscous diffusing liquid (glycerol + concentrated milk) is positioned in front of MO1 to randomize the illumination structure in both time and space. The incoming field $E_1$ is therefore randomly distributed over all the incoming modes and varies in time. Thus, considering $\left|t-t'\right|>100\ \rm ms$, the fields $E_1(t)$ and $E_1(t')$ are uncorrelated.

Measurement of the outgoing field $E_2$ is holographically performed. The two orthogonally polarized reference beams $R(p)$ (where $p=1,2$ is polarization) interfere with the outgoing fields $E_2(p)$, and the interference pattern  $I=\sum_{p=1,2}|R(p)+E_2(p)|^2$ is recorded on the CCD sensor (10 Hz, $1340\times1040$ pixels with $\Delta x=6.45\ \mu\rm m$ pitch).
This configuration makes it possible to calculate from $I$ the complex amplitudes $E_2(p)$ of the outgoing fields  along both polarizations $p=1,2$ directions by filtering, in the Fourier space, the desired +1 grating order i.e. $E_2(p)R^*(p)$ \cite{cuche2000spatial}.
%n order to reduce the influence of the zero grating order term $|R(p)|^2$ these calculation are made from

%

\begin{figure}
\begin{center}
  \includegraphics[width=2.25 cm]{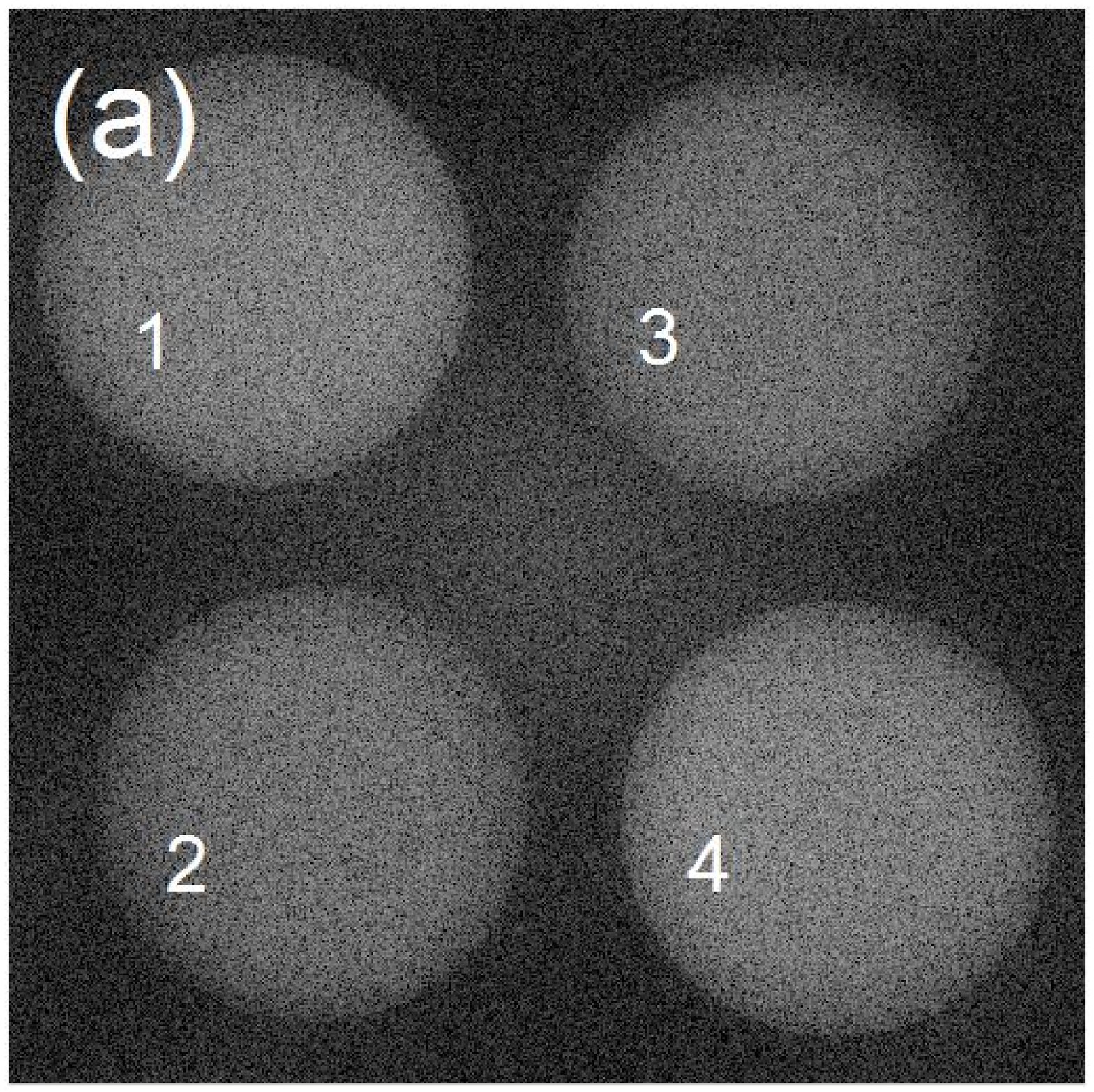}
   \includegraphics[width=2.25 cm]{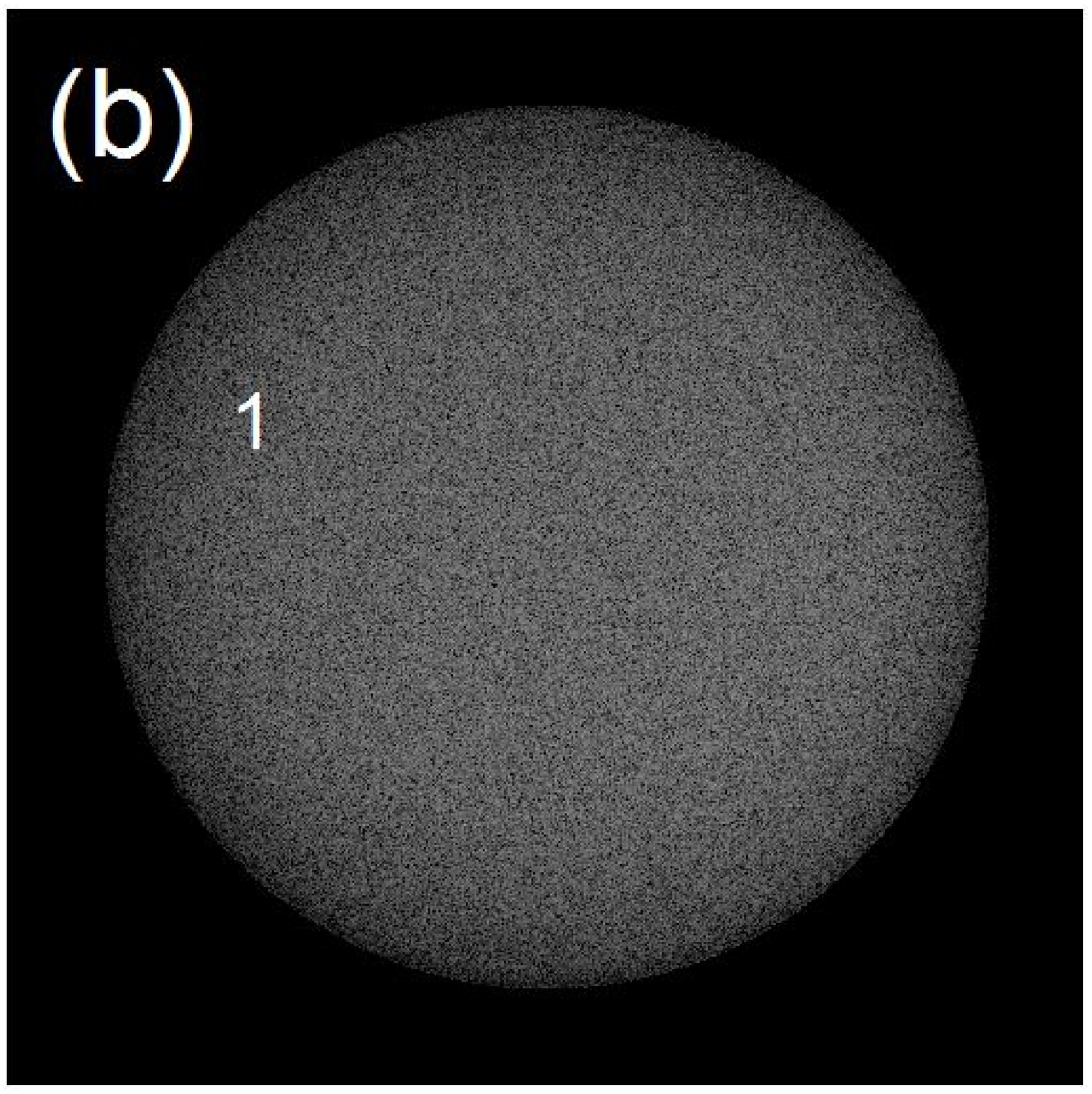}\\
  \includegraphics[width=2.25 cm]{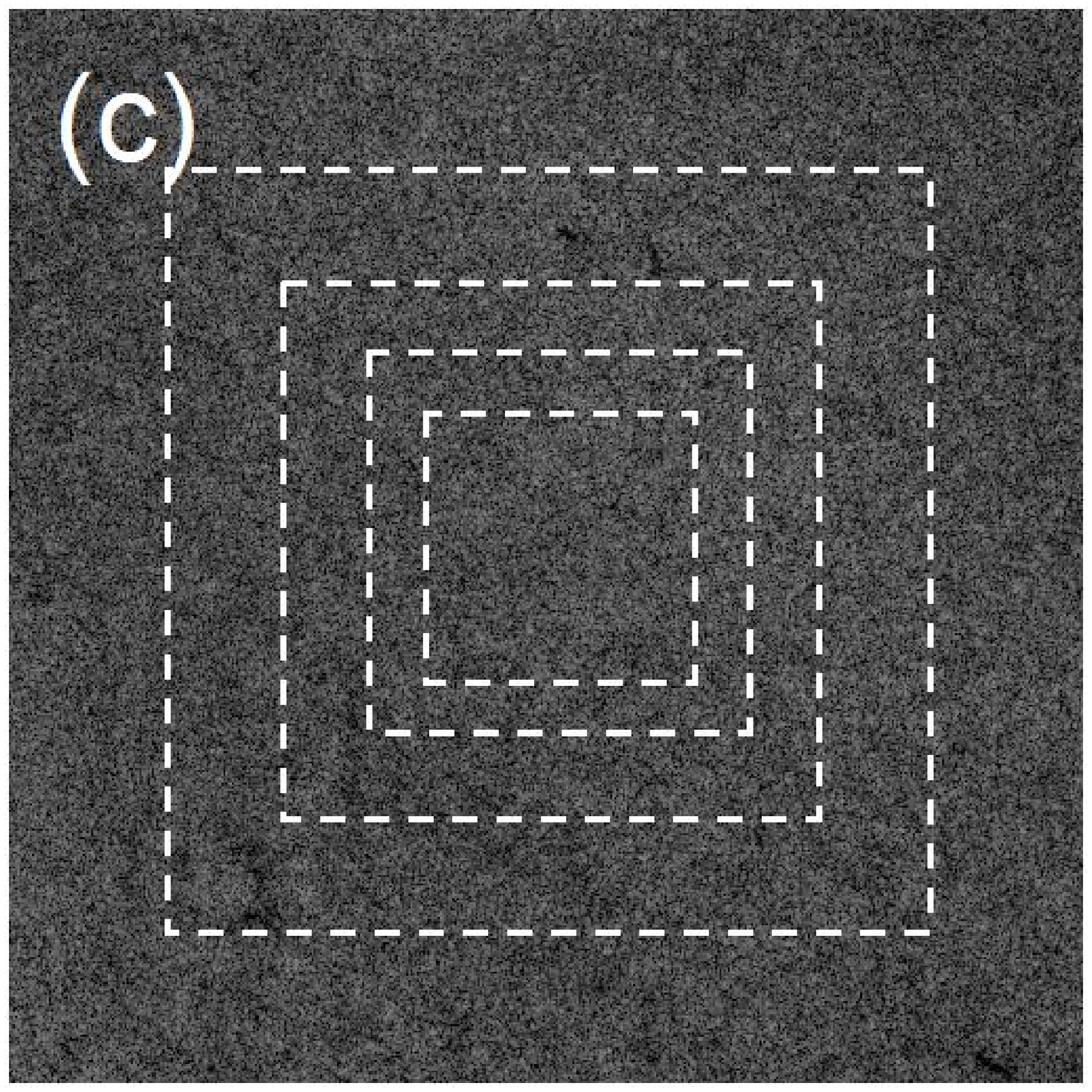}
   \includegraphics[width=2.25 cm]{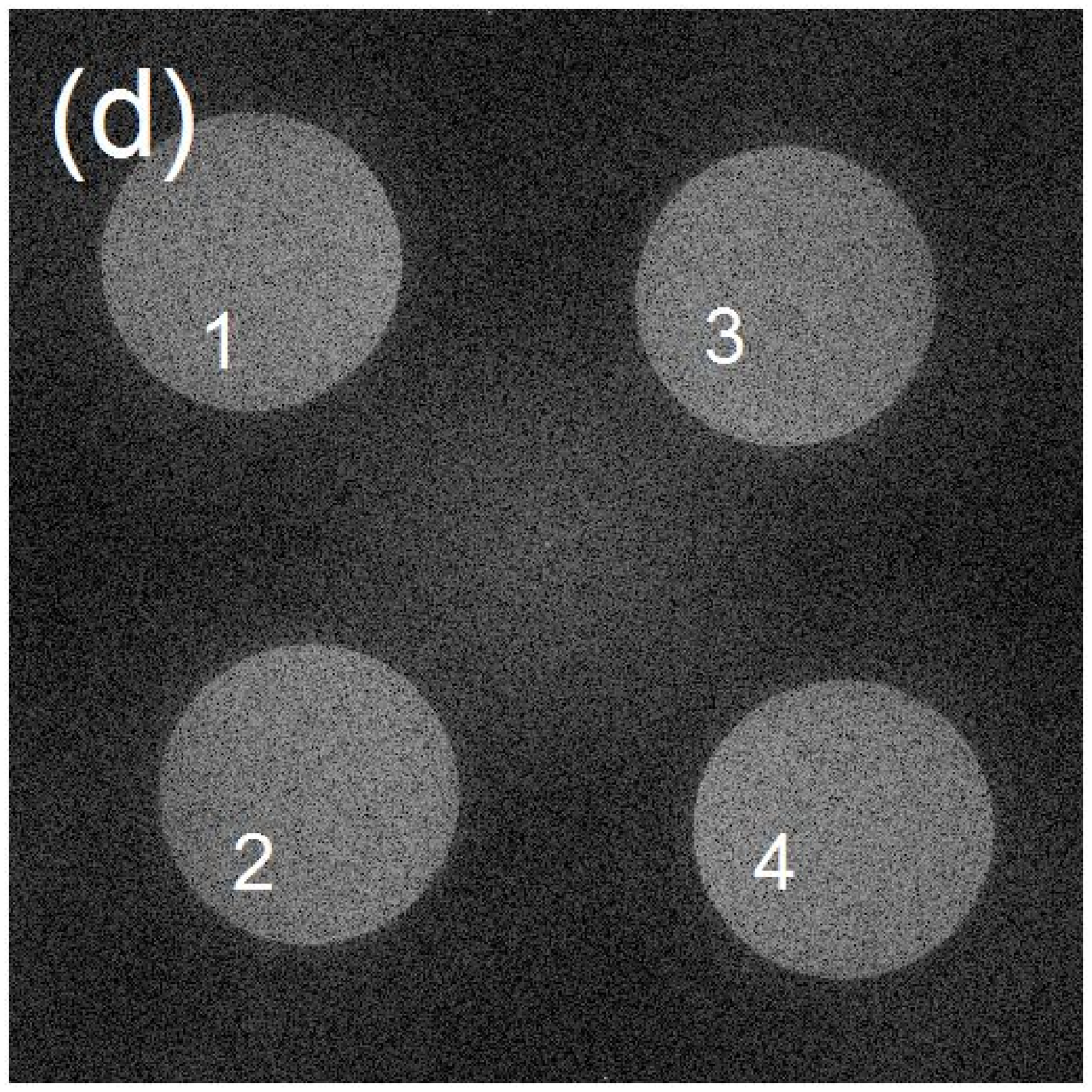}\\
 \caption{ $|{\tilde H}|^2$ (a,d);  $\times 2$ zoom of    $|{\tilde H}|^2$  near zone 1  (b);  $| E_2 R^*|^2$ calculated by Eq. \ref{Eq_E2R} for $p=1$ (c). In (c), the dashed squares  of size $M=$724, 512, 362  and 256 pixels illustrate the ability to select different zones of the sample   to calculate $\langle |C_2|^2\rangle$.  Images are obtained with ZnO sample (a-c), and  without (d).}\label{Fig_ima_pupil}
\end{center}
\end{figure}

The Fourier spacial filtering is illustrated by  Fig. \ref{Fig_ima_pupil}. The $1340\times1040$ holograms were  cropped to $1024\times 1024$  (the imaged area is therefore $L=77\ \mu\rm m$) and two frames holograms $H$ were calculated from successive frames:  $H(t_{2n})=I(t_{2n})- I(t_{2n+1})$. The  Fourier hologram  $\tilde{H}(k_x,k_y)={\rm FFT}\left[H\left(x,y\right)\right]$ (where ${\rm FFT}$ is the  Fast Fourier transform) is then calculated. $|\tilde{H}|^2$ is displayed on  Fig. \ref{Fig_ima_pupil}(a).  Four bright circular regions can be observed. They correspond to the reconstructed image of the MO2 pupil.   Reconstruction is   made with  +1 grating order $E_2(p)R^*(p)$ for $p=1$ (zone 1), and $p=2$  (zone 2), and to the -1 grating order $E^*_2(p)R(p)$ for $p=2$ (zone 3) and $p=1$ (zone 4).   Because the calculations are made with two frames hologram, the zero order terms $|R(p)|^2$ cancel, and  are  not visible on Fig. \ref{Fig_ima_pupil}(a).

Note that the angular tilt of the beam splitter BS  as well as the source point positions $F_1$ and $F_2$ (see Fig. \ref{Fig_Fig_setup}) have been chosen so that the four regions in Fig. \ref{Fig_ima_pupil}(a) do not overlap, and have sharp edges \cite{verrier2015holographic}.
From Fig. \ref{Fig_ima_pupil}(a), we have selected the desired +1 grating orders  $E_2R^*$ by cropping zone 1 and zone 2 (for $p=1$ and 2) and by taking the inverse Fourier transform of the cropped zones:
\begin{eqnarray}\label{Eq_E2R}
 E_2 R^*(x, y,p)={\rm FFT}^{-1} \rm{C}_p \left[ \tilde{H}(k_x,k_y,p)  \right]
\end{eqnarray}
where $\rm{C}_p$ is the crop operator for  polarisation $p$.

Since  $R(p)$ is roughly constant with the position $x,y$,  the correlations $C_2(t,t')$  can be then calculated by replacing  $E_2(x,y,p,t)$ by $[E_2 R^*](x,y,p,t)$ in Eq. \ref{eq1_corr_C_1_C_2_bis_bis}.
The statistical average $\langle |C_2|^2\rangle$ was obtained by first recording the sequences of 150 camera frames: $ I (t_0) ... I (t_ {149}) $, at times:  $t_n=n\Delta t$ and $\Delta t =100$ ms, yielding 75 hologram: $ H(t_0), H(t_2) ... H (t_ {148}) $, which were used to calculate
  $[E_2R^*](x,y,p,t_{2n})$ and $ C_2 (t_{2n}, t_{2n'}) $, and then by averaging    $ |C_2 (t_{2n}, t_{2n '})|^2 $ for all couple of times $ t_{2n}, t_{2n'} $  with $ | n-n '|> 5$  and $n,n'=0..74$.

Because of experimental defects,  $|R(p)|$ varies slightly with position. This  affects the calculation of   $C_2(t,t')$ and  $\langle |C_2|^2\rangle$. In order to account for this effect, we measured  $|R(x,y,p)|$  from our stack of holographic data and  we  calculated  $C_2(t,t')$ with $[E_2 R^*](x,y,p,t)/|R(x,y,p)|$. This  correction is  about 10\% for $\langle |C_2|^2\rangle$.

\section{Number of geometrical modes $N'_g$ }

%
%We conducted a control experiment without  ZnO sample and we calculated the correlation  $\langle |C'_2|^2 \rangle $ versus $z$ for $M=1024$. As seen on  Fig.\ref{fig__curves}  (b), $\langle |C'_2|^2 \rangle $  is much lower than with the sample, and does not depend on $ z $ position. Moreover, $\langle |C'_2|^2 \rangle \simeq$8.8e-6 is close to the correlation  that is expected without open  channels  $ \langle |C'_1|^2 \rangle =1/N'_g$=9.3e-6 (see solid grey line on  Fig.\ref{fig__curves}(b)). Note here that the number of geometrical mode $N'_g$ must be calculated by Eq.\ref{Eq_N_g} with  MO1 numerical aperture (i.e. NA=0.9).
%%
%Indeed, the signal collection angle is limited by MO1. %, and therefore the relevant  numerical aperture is that of MO1.
%This point is illustrated by

Figure \ref{Fig_ima_pupil} (d) displays the hologram $ |\tilde H|^2 $ we got without sample. $ |\tilde H|^2 $ exhibits four bright circular zones that are smaller in diameter  than with the sample (Fig.\ref{Fig_ima_pupil}(a)). These circles correspond to the MO1 pupil located in the plane P' that appears sharp in the  plane P, because MO1 and MO2 form an afocal optical system. There is thus no field out of the MO1 collection angle, and the number of geometrical mode $N'g$  must be calculated with MO1 numerical aperture NA=0.9.

We must notice that $N'_g$ is a little bit smaller than   the number of pixels of zones 1 and 2 in Fig. \ref{Fig_ima_pupil} (d). Similarly $N_g$ is a little bit smaller than the number pixels of zones 1 and 2 in Fig. \ref{Fig_ima_pupil} (a), but the difference is larger, because the brightness within the pupil decreases noticeably near the pupil edge.  This means that the reconstructed field within  the pupil is random from one pixel to the next.

We  used this property to calculate $N_g$ and $N'_g$.  We  measured the averaged intensities $\langle |{\tilde H|^2} \rangle$ for each pixels of zone 1 and 2, and we used this information to calculate by Monte Carlo the residual correlation $\langle |C_2|^2\rangle$ that is expected for pupils fields random in space and time. The number of mode  $N_g $ and $N'_g$ we got by this way, with and without sample, agree within a few per cent with $2\pi [\textrm{NA}]^2L^2/\lambda^2$ with NA=1.4 and 0.9.

\bibliographystyle{unsrt}
%\bibliography{biblio}